\documentclass[useAMS,usenatbib,usegraphicx,onecolumn]{mn2e}
\newcommand{\bvect}[1]{\mbox{\boldmath{$#1$}}}
\newcommand{\erf}{\mbox{$\mathrm{erf}$}}

\title[Protodiscs around Hot Magnetic Rotator Stars]{Protodiscs
around Hot Magnetic Rotator Stars}
\author[M. Maheswaran and J. P. Cassinelli]{M. Maheswaran$^1$ and J. P. Cassinelli$^2$
\thanks{Email:m.maheswaran@uwc.edu(MM); cassinelli@astro.wisc.edu(JPC)} \\ $^{1}$Dept. of Mathematics,
University of Wisconsin Marathon County, 518 S. 7th Avenue,
Wausau, WI 54401, USA \\
$^{2}$Dept. of Astronomy, University of Wisconsin-Madison, 475 N.
Charter St., Madison, WI 53706, USA}
\begin{document}
\date{Accepted ................  Received ...................}
\pagerange{\pageref{firstpage}--\pageref{lastpage}}
\pubyear{2008}

\maketitle

\label{firstpage}

\begin{abstract}
We develop equations and obtain solutions for the structure and evolution of a protodisc region that is initially formed with no radial motion and super-Keplerian rotation speed when wind material from a hot rotating star is channelled towards its equatorial plane by a dipole-type magnetic field. Its temperature is around $10^7$K because of shock heating and the inflow of wind material causes its equatorial density to increase with time. The centrifugal force and thermal pressure increase relative to the magnetic force and material escapes at its outer edge. The protodisc region of a uniformly rotating star has almost uniform rotation and will shrink radially unless some instability intervenes. In a star with angular velocity increasing along its surface towards the equator, the angular velocity of the protodisc region decreases radially outwards and magnetorotational instability (MRI) can occur within a few hours or days. Viscosity resulting from MRI will readjust the angular velocity distribution of the protodisc material and may assist in the formation of a quasi-steady disc. Thus, the centrifugal breakout found in numerical simulations for uniformly rotating stars does not imply that quasi-steady discs with slow outflow cannot form around magnetic rotator stars with solar-type differential rotation.
\end{abstract}


\section{Introduction}\label{sec-introduction}
We use the term \emph{protodisc} to refer to the disc region that is initially formed when wind material from a star is channelled by dipole-type magnetic fields towards the equatorial plane. The material in a protodisc region has no radial motion and rotates with the magnetic field. It is bounded above and below by shock boundaries. We develop models for the structure and evolution of protodiscs around magnetic rotator stars. We study protodiscs as separate entities while recognising that they may be precursors of quasi-steady discs, such as those of Be stars. We do not construct models of quasi-steady discs in this paper but indicate how protodiscs relate to models of Be star discs. For example, a protodisc region could initially be present in the inner part of the quasi-steady disc model with slow outflow studied by \citet{bro08}. Also, a more detailed study of protodiscs is important to verify how centrifugal outflow in protodiscs of rapidly rotating stars \citep[e.g.,][]{mah03,udd06} affects the formation of quasi-steady discs around magnetic rotator stars. In this paper we consider protodiscs around stellar models with both uniform and differential rotation. We study how their radial extents change with time and we determine the conditions under which MRI will occur to affect the angular velocity distribution and the radial outflow in the disc region.

There has been a tremendous increase in observations of emission line B stars, including polarisation, interferometry and magnetic diagnostics, but the question of how discs with near Keplerian rotation speeds are formed around them is has not been fully answered. There can be other mechanisms that are responsible for disc formation in different stars depending on their properties. For example, \citet{lee91} have proposed a decretion disc model for Be stars with near maximal rotation speeds. There is no magnetic force in
their model and angular momentum is transported by viscous forces. Be stars are known to rotate rapidly \citep[e.g., see][]{por03} and the presence of significant magnetic fields in OB stars has been reported by several authors \citep[e.g.,][]{don06, nei07,
sch07}. The concept of a Wind Compressed Disc was initially developed by \citet{bjo93}. Subsequently, \citet{cas02} proposed a Magnetically Torqued Disk(MTD) model in which a sufficiently strong magnetic field channels a flow of wind material towards the
equatorial plane to form a disc and provides the torque required by the disc to maintain quasi-Keplerian rotation speeds. \citet{mah03, mah05} developed the Magnetic Rotator Wind
Disk(MRWD) model in which Keplerian discs may be formed with stellar magnetic fields that do not have to be as strong as those required in the MTD model. \citet{bro04} presented an amended version of the MTD model to include stellar gravity darkening and \citet{brc05} suggested that this should be further modified to include the effect of gravity arising from finite thickness of the disc. \citet{bro08} have studied the properties of discs with slow radial outflow and address criticisms of the MTD model.

In the MRWD model for rotating stars with winds, when magnetic field lines that emerge from the stellar surface loop back to the star, a transient disc that has no meridional motion and corotates with the star is initially formed. After the local density in the disc has increased to a value where the centrifugal force exceeds the sum of gravity and the magnetic force, material diffuses radially outwards while stretching out the magnetic field lines. \citet{mah03} suggested that such material can diffuse into quasi-steady discs and that the presence of viscosity during this stage would enable the formation of Keplerian discs of larger radial extent. On the other hand, numerical simulations by \citet{udd06, udd08} for the flow of wind material along dipole magnetic field lines display centrifugal breakout in the form of transient bursts and they concluded that MTD type models were not
appropriate to explain disc formation. However, there are problems with their simulations: (a) As pointed out by \citet{bro08} they did not follow changes for a sufficiently long duration of time. (b) The simulations are essentially two dimensional. (c) Numerical MHD simulations do not always provide accurate results because discretization of the MHD induction equation leads to the introduction of artificial magnetic diffusivity that is several orders of magnitude larger than the real magnetic diffusivity in discs. This can yield spurious results for magnetic reconnection that may not be realistic. (d) Their models fail to consider differentially rotating stars and, therefore, do not obtain differentially rotating disc regions that are subject to the onset of MRI. The occurrence of MRI can significantly affect the rotation speed in a disc region. We carry out an analytical study and obtain time-dependent solutions for uniform and differentially rotating stars. We find that when the stellar angular velocity increases along the surface towards the equator there is a broad range of stellar parameters for which MRI occurs in a sufficiently large part of the protodisc region.

We consider a stellar model in which the magnetic field is dipole-type over some region of the stellar surface, so that field lines leaving the stellar surface pass through the equatorial plane and return to the star. Wind material flowing along such lines is channelled into a dense equatorial protodisc region. Since the flow of supersonic wind material into this region from above and below are in opposite directions, this material will cross shock surfaces that form the upper and lower boundaries of the disc region. We refer to the region between the sonic surface of the wind and the shock boundary of the disc as the \emph{wind zone}. Thus, we need not be concerned about gas pressure effects on the wind flow, which is treated in an empirical fashion in that the velocity is chosen to obey a beta law and the density is assumed to follow from a global mass conservation relation. In section \ref{sec-windzone}, we set up the relevant properties of axially symmetric wind zones that enable us to determine values of physical quantities at points on the wind side of the shock boundary from known values at the stellar surface. We develop formalism to account for the presence of differential rotation along the region of the stellar surface from which wind material flows into the disc region. The material in a protodisc region has no radial motion and rotates with magnetic field lines because the
centrifugal and pressure gradient forces that are directed outwards are balanced by gravity and magnetic force. In section \ref{sec-proto-region} we describe the general concept of a
protodisc, which is a more generalised version of the pre-Keplerian disc developed in the MRWD model. Also, we describe the physical processes that are involved in their formation and evolution. In section \ref{sec-shock-proto-jump-conditions}, we develop MHD generalisations of the Rankine-Hugoniot jump conditions that are appropriate for shock boundaries of protodiscs so that we can compute changes in values of quantities across
shock surfaces. We take into account the vertical motion of shock surfaces caused by continuous filling up of a protodisc region by wind material.

In section \ref{sec-disc-proto-eqns}, we develop the equations that govern the structure of the models of protodiscs that we consider. We explicitly include the effect of the component of stellar gravity perpendicular to discs and, because of shock heating of the disc region, we include the vertical and radial components of the thermal pressure gradient in the respective momentum equations for protodiscs. Even though the heights of protodiscs may be relatively small, the presence of a vertical component of stellar gravity leads to a vertical gradient of pressure and density. The equatorial density of material in the protodisc increases with time as the disc fills up. This has the effect of increasing the
centrifugal force in relation to the magnetic force as well as increasing the thermal pressure in relation to magnetic pressure. The first of these effects tends to cause material at the outer edge of a protodisc to move radially outwards whereas the latter
leads to the onset of MRI within the protodisc. To determine the relative importance of these competing effects, in section \ref{sec-onset-mri-protodiscs} we derive criteria that influence the evolution of protodiscs. Also, we discuss the conditions under which the shrinking of a protodisc caused by centrifugal outflow is halted by the onset of MRI. This leads to viscous readjustment of the rotation speed of the disc region. In section
\ref{sec-disc-proto-applics-results}, we apply our equations and jump conditions to protodiscs of different stellar models and present the results of our computations. In section \ref{sec-evln-protodisc-after-mri}, we discuss important factors that affect the evolution of protodisc material after the onset of MRI and refer to some relevant quasi-steady models. Our conclusions regarding the formation and evolution of protodiscs around rotating magnetic stars with winds are summarised in section \ref{sec-discussion}.
An important objective of this paper is to relate the properties of protodisc regions to values of quantities at the stellar surface. Therefore, we include all the equations that are necessary to solve for the different physical quantities in the wind zone, at the shock surface and in the protodisc region.

\section{Axially Symmetric Wind Zone}\label{sec-windzone}
In this section we present the equations that are used to determine the values of different physical quantities at points on the wind side of the shock boundary in terms of their values at the stellar surface. We assume that the wind zone is in a steady state and is isothermal. The mass, luminosity, radius and effective temperature of the star are $M, L, R$ and $T_{eff}$ respectively. We use an inertial frame of reference with spherical
coordinates $(r, \theta, \phi)$ in the wind zone and cylindrical coordinates $(\varpi, \phi, z)$ in the disc region. We assume that the star, wind and magnetic field are symmetrical about the axis of rotation. Let the angular velocity at any point $P$ on the stellar surface be $\omega$, which may vary with latitude along the stellar surface. We show in section \ref{sec-proto-region} that stellar models with appropriate differential rotation will have protodiscs that satisfy a condition that is necessary for the
occurrence of MRI. All the equations that we set up for the wind zone are valid even when $\omega$ varies along the stellar surface provided that the model is axially symmetric. We define an associated rotation rate parameter ${\mathcal S}_\star$ for the point $P$ by the equation $\omega = [GM/R^3]^{1/2} \, {\mathcal S}_\star$. If the star has uniform
rotation, then $\omega$ is constant for the star and ${\mathcal S}_\star$ gives the ratio of the equatorial rotation speed of the star to the Keplerian rotation speed. Otherwise, ${\mathcal S}_\star$ will vary with $\omega$ along the stellar surface. Let $\rho$ and $\bvect{B}$ denote density and magnetic field, respectively. We use $\bvect{v}$ and $\bvect{u}$, respectively, to denote velocities in the wind zone and the disc region. Subscript $m$ denotes values of meridional components. Then, $v_m=\left(v_r^2+v_\theta^2\right)^{1/2}$ and $B_m=\left(B_r^2+B_\theta^2\right)^{1/2}$. Let $Q$ be a point of impact of wind material on the shock boundary. Subscripts $W$ and
$D$, respectively, denote values of quantities on the wind side and the disc side at the point $Q$. Let the meridional magnetic field line that emerges from the point $P$ on the star pass through the point $Q$ on the shock surface.

Equations for the wind zone were set up in \citet{mah03} using the steady state equations given by \citet{mes68}. Initial values of different quantities in the wind zone are specified at the sonic surface and are denoted using subscript $\circ$. The MHD induction
gives $\rm{curl}(\bvect{v \times B})= \bvect{0}$, which implies that the meridional vectors $\bvect{v}_m$ and $\bvect{B}_m$ must be parallel to each other. Hence, the streamlines of flow in a meridional plane in the wind zone coincide with magnetic field lines. This, together with the mass and magnetic flux conservation conditions $\textrm{div}(\rho \bvect{v})=0$ and $\textrm{div} \, \bvect{B}=0$ require that
\begin{equation}\label{eqn-wind-rho-vm-bm} \frac{\rho \, v_m}{B_m} = \frac{\rho_\circ \, v_{m, \circ}}{B_{m, \circ}} =  \xi \, ,
\end{equation}
where $\xi$ is constant along meridional streamlines and represents the ratio of mass flux to magnetic flux. At the sonic surface, $v_{m, \circ}$ will be equal to the isothermal sound speed. In general, $r_\circ$ and the density $\rho_\circ$ at the sonic surface will be functions of $\theta$. In our models, we take the sonic surface to be approximately spherical so that $r_\circ$ is constant. If the stellar mass loss rate $\dot{M}$, is
known only as an average value for the star rather than as a function of $\theta$, we can write
\begin{equation}\label{eqn-wind-initial-condn-rho-vm}
4\pi r_\circ^2\rho_\circ \, v_{m, \circ} = \dot{M} \, ,
\end{equation}
where  $\rho_\circ$ is constant and represents the average density over the sonic surface. The azimuthal components of wind velocity and magnetic field satisfy the condition
\begin{equation}\label{eqn-wind-v-phi-omega}
v_\phi = \omega \varpi + v_m B_{\phi}/B_{m} \, .
\end{equation}
Equations (\ref{eqn-wind-rho-vm-bm}) and
(\ref{eqn-wind-v-phi-omega}) allow us to visualise the magnetic field line that is anchored at the point $P$ in the star to be rotating in the wind zone with the same angular velocity $\omega$ as  $P$. The azimuthal field $B_\phi$ is negative in the northern hemisphere of the wind zone so that field lines curve in the direction opposite to that of rotation. Since wind material flows along field lines, the rotation speed of the wind material is less than the corotation speed as verified in equation (\ref{eqn-wind-v-phi-omega}). In our stellar models, the material flowing into the disc region encounter the shock boundary before it could reach the Alfv\'en point. The azimuthal torque equation for the wind can be written in the form
\begin{equation}\label{eqn-wind-azimuthal}
\varpi \left(v_\phi - \frac{B_{m, \circ}\, B_\phi}{4 \pi
\rho_\circ v_{m, \circ}}\right) \, = \, \mathcal{L},
\end{equation}
where $\mathcal{L}$ is constant along a streamline and may be interpreted as total angular momentum per unit mass. Each field line in the wind zone is associated with a characteristic density $\rho_{A}$ and a characteristic distance $\varpi_{A}$ from the
rotation axis. These are given by
\begin{equation}\label{eqn-wind-rho-c}
\rho_{A} \, =  \, \frac{4 \pi \rho_\circ^2 v_{m, \circ}^2}{B_{m,
\circ}^2} \, = \, \frac{{\dot M}^2}{4 \pi r_\circ^4 \, B_{m,
\circ}^2}
\end{equation}
and
\begin{equation}\label{eqn-wind-ell}
\varpi_{A}^2 = \frac{\mathcal{L}}{\omega } =
\left(\frac{\varpi_\circ v_{\phi, \circ}}{\omega} \right) \left(1
- \frac{B_{m, \circ} \, B_{\phi, \circ}}{4 \pi \rho_\circ \, v_{m,
\circ} \, v_{\phi, \circ}} \right)   \, .
\end{equation}
In general, the values of $\rho_{A}$ and $\varpi_{A}$ will vary with the latitude of the point $P$ on the stellar surface from which the corresponding field line emerges. Note that a streamline that does not enter the disc region but travels with the wind to a
large distance will have an Alfv\'en point where $\rho = \rho_A$ and $\varpi = \varpi_A$ . However, for streamlines that enter the shock boundary of a protodisc region, there is no point in the wind zone where $\rho$ and $\varpi$ are equal to these Alfv\'en values but we define them as streamline constants because we can use them to compute the rotation speed and azimuthal component of the magnetic field at any point in the wind zone. We solve
equations (\ref{eqn-wind-v-phi-omega}) and (\ref{eqn-wind-azimuthal}) to obtain the values of the $\phi$-components of the wind velocity and magnetic field. Then, using equations (\ref{eqn-wind-rho-c}) and (\ref{eqn-wind-ell}) we obtain
\begin{equation}\label{eqn-wind-v-phi}
v_{\phi, W} = v_{\phi, \circ}
\left(\frac{\varpi}{\varpi_\circ}\right)
\left\{\frac{\left[1-\left(\rho_{A} \varpi_{A}^2\right)/\left(\rho
\varpi^2\right)\right] \left(1-{\rho_{A}}/{\rho_\circ}\right)}
{\left[1-\left(\rho_{A} \varpi_{A}^2\right)/\left(\rho_\circ
\varpi_\circ^2\right)\right]
\left(1-{\rho_{A}}/{\rho}\right)}\right\} \,
\end{equation}
and
\begin{equation}\label{eqn-wind-B-phi}
B_{\phi, W} = B_{\phi, \circ}
\left(\frac{\varpi}{\varpi_\circ}\right)
\left\{\frac{\left(1-\varpi_{A}^2/\varpi^2\right)
\left(1-{\rho_{A}}/{\rho_\circ}\right)} {\left(1- \varpi_{A}^2/
\varpi_\circ^2\right) \left(1-{\rho_{A}}/{\rho}\right)}\right\} \,
,
\end{equation}

We express the meridional magnetic field in the wind zone in the form for dipole-type fields developed in \citet{mah03}. We have $B_r = 2C\left({r_\circ}/{r}\right)^b \cos \theta$ and $ B_\theta = (b-2) C\left({r_\circ}/{r}\right)^b \sin \theta$, where $b$ is a
constant such that $2 < b \leq 3$. We should point out that $b = 3$ represents a strict dipole field and that the the meridional component of the magnetic force will be zero in this case. However, for $2 < b < 3$ there is a nonzero meridional component of the magnetic force. As was pointed in the earlier paper, even if $b = 3$ at the stellar surface its value will be less than 3 in the wind zone because the field lines will be stretched out to some extent by the wind material. We obtain the meridional component of the magnetic field in the wind zone as
\begin{equation}\label{eqn-wind-Bm}
B_m = C \left(\frac{r_\circ}{r}\right)^b \,
\left[4-b(4-b)\sin^2\theta \right]^{1/2} \, .
\end{equation}
In the case of disc regions with relatively small height, we have $\sin\theta \approx 1$  at points along the shock boundary. Then, equation (\ref{eqn-wind-Bm}) gives an approximation for $B_{m, W}$ in the form
\begin{equation}\label{eqn-wind-BmW}
{B_{m, W}} \approx (b-2) \, C \left({\frac{r_\circ}{r}}\right)^b
\, .
\end{equation}
To obtain the meridional speed $v_{m, W}$ of the wind material as it enters the shock boundary, we use the beta velocity law for line driven winds given by
\begin{equation}\label{eqn-wind-beta-law}
v_{m, W} = V_\infty \left(1-\frac{R}{r}\right)^{\beta_{vel}} \, ,
\end{equation}
where $\beta_{vel}$ is a constant that is in the range $0.5 < \beta_{vel} < 2$ \citep[see e.g.,][]{lam99}. Also, $V_\infty$ is a terminal speed that is constant for each streamline but may be different for different streamlines. In our computations we take $V_\infty$ to be the same for all streamlines and equal to the velocity at infinity of the wind. The density $\rho_W$ of the wind at the shock surface is obtained in terms of the density at the initial surface given by equation (\ref{eqn-wind-initial-condn-rho-vm}) by using the mass and magnetic flux continuity equation (\ref{eqn-wind-rho-vm-bm}) together with the velocity law (\ref{eqn-wind-beta-law}). In stellar models where the ratio $\Gamma$ of the stellar luminosity to the Eddington luminosity becomes important, we replace $M$ by
$M(1-\Gamma)$ in the appropriate equations.

We introduce the variable $x = \varpi/R$ to denote the non-dimensional distance $x$ of the point $X$ on the equatorial plane, whose distance from the centre of the star is $\varpi$. When a subscript is added to $X$ we add the same subscript to the corresponding $x$. The wind material that is channelled along flux tubes of a magnetic field crosses a shock surface and moves into an equatorial region between the stellar equator and a point
$X_{lim}$, which is the outermost point on the equatorial plane to which the magnetic field is able to channel the flow of wind material. Let $X_{kep}$ be the innermost point on the equatorial plane where local Keplerian rotation speed equals the corotation speed. Then, $x_{kep} = \mathcal{S}_\star^{-2/3}$, where $\mathcal{S}_\star$ is the rotation rate of the star at the point $P_{kep}$ on the stellar surface that is linked to $X_{kep}$ by a
magnetic field line \citep{cas02}. The conditions under which the stellar magnetic field will be able to channel the wind into a disc region and relevant parameters are discussed in detail in sections 3, 4 and 10 through 13 in \citet{mah03}. In particular, the value of the meridional Alfv\'en parameter $A_{m} = B_{m}/(\sqrt{4 \pi \rho} \, v_m)$ must be larger than unity at all points along a streamline from the star to the shock surface. Hence, the strength of the meridional magnetic field at the initial surface must be larger than a minimum value given by $B_{m, 0, min}$ for the formation of protodiscs. Values of $B_{m, 0, min}$ are approximately 1G for B5 or B9 stars, 10G for B2 stars, 60G to 80G for B0 stars and of the order of 100G for O type stars. Our interest is in stars with magnetic fields that are only slightly stronger than these minimum values so that they are able to channel wind flow into a protodisc region. In this paper we consider models of O and B type stars shown in table \ref{table-star-properties} with rapid rotation such that $\mathcal{S}_\star \ga 0.5$.

\section{Protodisc Regions} \label{sec-proto-region}
\begin{figure}
\includegraphics[width=120mm]{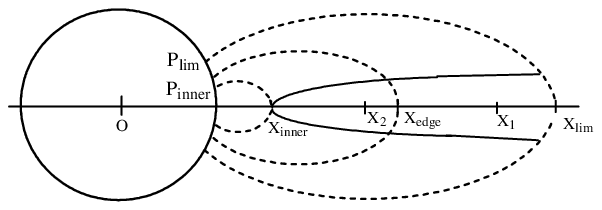}
\caption{Schematic figure of a meridional cross-section of a protodisc region of the star. Dashed lines represent the coincident streamlines and magnetic field lines.} \label{fig1}
\end{figure}

Here, we describe the physical properties of protodiscs and the processes involved in their formation and evolution. A magnetic field of sufficient strength in a hot rotating star channels the flow of wind material from the star into an equatorial region. Furthermore, in some part of this region, the field lines initially loop back to the star and the magnetic force is strong enough to provide support against centrifugal force so that the material accumulates to form a protodisc with  no meridional motion. It is bounded above and below by shock surfaces as a result of the flow of wind material towards the equatorial plane from opposite hemispheres. Figure \ref{fig1} gives a schematic meridional cross-section of a protodisc region in an axially symmetric star. Let $\delta$ denote the angle that the tangent line to the shock surface at the point $Q$ makes with the equatorial plane. We use subscript $n$ to denote the normal component of a vector at the shock boundary in the direction from the wind zone into the disc region. Let $h$ be the height of the point $Q$ above the
equatorial plane. We assume that the material in the disc region satisfies the ideal gas law $p = \mathcal{R} \rho T/\mu$, where $\mathcal{R}$ is the gas constant and $\mu$ is the mean molecular mass. For convenience, we use the substitution $p = \rho a^2$, where $a^2 =\mathcal{R} T/\mu$. Note that the adiabatic sound speed is $a \gamma^{1/2}$, where $\gamma$ is the usual ratio of specific heats.

Let $X_{inner}$ and $X_{edge}$ shown in Figure \ref{fig1} be the innermost and outermost points of the region in the equatorial plane where the magnetic force is strong enough to prevent radial motion against the effects of the centrifugal, gravitational and thermal pressure forces. The point $X_{inner}$ will be located very close to the point $X_{kep}$ and its position will be influenced by the radial components of the magnetic and thermal pressure force in addition to gravity and centrifugal force. A protodisc will form in the region between the points $X_{inner}$ and $X_{edge}$. Wind material arriving between the stellar equator and $X_{inner}$ will tend to fall back towards the stellar surface because the sum of gravity and magnetic force dominates the sum of centrifugal and pressure forces. Material arriving between $X_{edge}$ and $X_{lim}$ will flow outwards because the centrifugal force dominates in this region. Meridional magnetic field lines emerging from the points $P_{inner}$ and $P_{lim}$ on the stellar surface are connected to the points
$X_{inner}$ and $X_{lim}$, respectively, on the equatorial plane.

The protodisc region has super-Keplerian rotation speed. As this region fills up, the density at points along the equatorial plane increases and the centrifugal force increases relative to the sum of gravity and magnetic force. While the location of $X_{inner}$
remains fixed, centrifugal outflow causes the outer edge of a protodisc $X_{edge}$ to move towards $X_{inner}$ as time $t$ increases, provided that factors such as MRI do not intervene. At the same time, the continuing impact of wind material on the shock boundary causes the protodisc to heat up. The thermal pressure can rise to values where it exceeds the local magnetic pressure and the ratio $\beta_{mag} = 8 \pi \rho a^2/B^2$ can become larger than unity. Let the angular velocity of the disc material at a point $X$ on the equatorial plane be $\Omega$ so that $u_{\phi, X} = \Omega \varpi$. There will be an onset of MRI when the two conditions
\begin{equation}\label{eqn-disc-proto-MRI-conditions}
\beta_{mag} > 1 \hspace{12pt} \textrm{and} \hspace{12pt}
\frac{\partial \, \Omega} {\partial\varpi} < 0
\end{equation}
are satisfied \citep{bal91}. Then, viscosity will become an important factor in the rotation of the disc region and the magnetic field will be affected by turbulence resulting from MRI.

In general, the angular velocity $\omega$ of the region of the stellar surface from which material flows to a protodisc region will vary with latitude. From section \ref{sec-windzone}, we know that $\omega$ is constant along a meridional magnetic field line in the wind zone. In our model of a protodisc, each magnetic field line from the stellar surface that enters the protodisc region is continuous across the protodisc and returns to the stellar surface. In the case of stars that have differential rotation in which $\omega$ increases along the stellar surface from the point $P_{lim}$ to the point $P_{inner}$, if the shock boundary of the protodisc is stationary, the protodisc rotation theorem established in \citet{mah03} shows that there will be a corresponding decrease in the angular velocity $\Omega$ of the protodisc region from $X_{inner}$ to $X_{edge}$. That is, the second of the two necessary conditions (\ref{eqn-disc-proto-MRI-conditions}), ${\partial \, \Omega}/{\partial\varpi} < 0$, for the onset of MRI in protodiscs
is satisfied when the angular velocity $\omega$ increases along the stellar surface towards the stellar equator and the shock boundary of the protodisc is stationary. In section
\ref{sec-onset-mri-protodiscs}, we consider the case where the shock boundary of the protodisc is not stationary but moves so that its height changes with time. Here too, results of numerical computations presented in section \ref{sec-disc-proto-applics-results} show that for appropriate models of differential rotation of the star where $\omega$ increases towards the equator, the condition ${\partial \, \Omega}/{\partial\varpi} < 0$ is satisfied in the protodisc. The first condition for the onset of MRI will be satisfied when the increase in density and shock heating of the protodisc region cause  $\beta_{mag}$ to become larger than unity. \citet{bal91} have also shown that the rapidly growing modes
increase exponentially on a time scale of order $4/(3 \, \Omega)$ seconds. The rotation time scale for the protodisc region is $t_{rot} = 2 \pi/\Omega$ seconds. Thus, MRI will occur on a time scale that is even shorter than the rotation time scale when the necessary conditions for the onset of MRI are satisfied. Although we focus on the onset of MRI, it is possible that other types of MHD instability could occur in protodiscs. Our interest is in
identifying at least one mechanism that will prevent a protodisc from shrinking completely and redistribute its angular momentum so that a quasi-steady disc may be formed.

We consider the protodisc phase of a disc to end at the time that MRI occurs in that region. In an axially symmetric protodisc, the meridional component of the velocity is negligible and we have $\bvect{u} = (0, u_\phi, 0)$ in cylindrical coordinates. The magnetic field is given by $\bvect{B} = (B_\varpi, B_\phi, B_z)$. When $h/\varpi$ is relatively small, the slope $\tan\delta$ of the shock surface is also small and the normal
component wind velocity $v_{n, W}$ is a function of $\varpi$ only along the shock boundary. Then, because the ram pressure from inflowing wind material on the shock boundary remains steady, the temperature $T_D$ of the protodisc at a point $Q$ of the shock will also be function of $\varpi$ only and does not change as $h$ increases. Thus, if the radiative cooling time of the protodisc is much longer than the flow time of wind material into the shock surface, the temperature $T$ in the protodisc will have the same value along a vertical line because there is no meridional flow of material in a protodisc region.
That is, we can take $T$ to be independent of $z$ so that $T = T(\varpi) = T_D(\varpi)$ and $\partial a/\partial z = 0$. In models where $h/\varpi$ becomes a large fraction before the onset of MRI it will be necessary to solve for $T$ as a function of $\varpi$ and $z$.

\section{Jump Conditions across Protodisc Shock Boundaries}\label{sec-shock-proto-jump-conditions}
In this section we set up the equations that are necessary to determine values of quantities at points on the disc side of the shock surface from the values on the wind side that we obtain in section \ref{sec-windzone}. As material flows from the wind zone
into a protodisc region, the height $h$ of the shock surface forming the boundary of the protodisc increases. Let $-U_{s}$ denote the normal component of the velocity of the point $Q$ on the shock surface in the direction of $n$. To solve for the properties of a protodisc we need the values of $\rho_D$, $p_D$, $U_{s}$, $u_{\phi, D}$, $B_{n, D}$ and $B_{\phi, D}$ at points on the disc side of the shock boundary. These are obtained by solving the six Rankine-Hugoniot jump conditions (\ref{eqn-shock-disc-proto-mass-continuity}) through (\ref{eqn-shock-disc-proto-energy}).

The continuity of mass across the shock boundary gives
\begin{equation}\label{eqn-shock-disc-proto-mass-continuity}
\rho_W \left(v_{n, W} + U_{s}\right)\, = \, \rho_D \, U_{s} \, .
\end{equation}
Because the normal component of the magnetic field and the tangential component of $\bvect{v \times B}$  must be continuous across the shock boundary, we have
\begin{equation}\label{eqn-shock-disc-proto-Bn}
B_{n, W} = B_{n, D}
\end{equation}
and
\begin{equation}\label{eqn-shock-disc-proto-vxB-tangential}
\left(v_{n, W} + U_s\right) B_{\phi, W} - v_{\phi, W}B_{n, W} =
U_s B_{\phi, D} - u_{\phi, D} B_{n, D} \, .
\end{equation}
Continuity of the normal component of momentum gives the jump condition
\begin{equation}\label{eqn-shock-disc-proto-mom-normal}
p_W + \rho_W  \left(v_{n, W} + U_s\right)^2 + \frac{B_W^2}{8\pi} =
p_D + \rho_D U_s^2 + \frac{B_D^2}{8\pi}
\end{equation}
and the equation for the continuity of the $\phi$-component of momentum is
\begin{equation}\label{eqn-shock-disc-proto-mom-phi}
\rho_W  \left(v_{n, W} + U_s\right) v_{\phi, W} - \frac{B_{n,
W}B_{\phi, W}}{4\pi} = \rho_D U_s u_{\phi, D} - \frac{B_{n,
D}B_{\phi, D}}{4\pi} \, .
\end{equation}
The jump condition for energy across the shock surface is obtained by using the condition that the sum of the kinetic energy, internal energy, thermal energy and electromagnetic energy must be conserved. Then,
\begin{equation}
\begin{array}{c}\label{eqn-shock-disc-proto-energy}
\rho_W (v_{n, W} + U_s)\left[\frac{(v_{m, W} + U_s)^2 + v_{\phi,
W}^2}{2} + \frac{\gamma \, p_W}{(\gamma -1) \, \rho_W} +
\frac{B_{\phi, W}^2}{4\pi \rho_W}\right]
- \frac{v_{\phi, W} B_{n, W} B_{\phi, W}}{4 \pi} \\
= \rho_D U_s \left[\frac{U_s^2 + u_{\phi, D}^2}{2} + \frac{\gamma
\, p_D}{(\gamma -1) \, \rho_D} + \frac{B_{\phi, D}^2}{4\pi
\rho_D}\right] - \frac{u_{\phi, D} B_{n, D} B_{\phi, D}}{4 \pi}
\end{array}
\end{equation}
for the flow of energy per unit area normally across the shock surface. When both sides of this equation are expanded to remove the brackets, we obtain four terms on each side. The first term represents the flow of kinetic energy of the material and the second term is the sum of the flows of internal energy and thermal energy. The sum of the third and fourth terms gives the normal component of the Poynting vector that represents the flux of electromagnetic energy of the plasma. The Rankine-Hugoniot equations that we have set up above have general validity for protodiscs. However, in the case of the protodisc models that we consider here, the changes in values of $u_\phi$ and $B_\phi$ across the shock surface are not large. Thus, we find that the influx of mass, momentum and kinetic energy of the wind material go mainly into increasing the the density and temperature of the
protodisc.

Equations (\ref{eqn-shock-disc-proto-vxB-tangential}) and (\ref{eqn-shock-disc-proto-mom-phi}) give the azimuthal components of velocity and magnetic field on the disc side of the shock surface as
\begin{equation}\label{eqn-shock-disc-proto-uphiD}
u_{\phi, D} = v_{\phi, W} - \frac{v_{n, W} \, B_{n, W} \, B_{\phi,
W}}{4 \pi \rho_D \left(\mathcal{A}_{n, D}^2 - U_s^2\right)}
\end{equation}
and
\begin{equation}\label{eqn-shock-disc-proto-BphiD}
B_{\phi, D} = B_{\phi, W} - \frac{U_s  \, v_{n, W} \, B_{\phi,
W}}{\left(\mathcal{A}_{n, D}^2 - U_s^2\right)} \, ,
\end{equation}
where $\mathcal{A}_{n, D} = B_{n, D}/\sqrt{4 \pi \rho_D}$ is the normal component of the Alfv\'en velocity at a point on the disc side of the shock boundary. In our protodisc models we have $\mathcal{A}_{n, D}^2$ larger than $U_s^2$. This is a consequence of the condition that the stellar magnetic field is able to channel the flow of wind material into the disc region. Since $B_{\phi, W} < 0$, equations (\ref{eqn-shock-disc-proto-uphiD})
and (\ref{eqn-shock-disc-proto-BphiD}) show that the rotation speed and azimuthal component of the magnetic field increase across the shock boundary. In our protodiscs we find that
$u_{\phi, D}$ is at most 2\% larger than the corotation speed $\omega \varpi$. In protodiscs where the shock surface is stationary, that is $U_s = 0$, we find that $u_{\phi, D} = \omega \varpi$, which is consistent with the protodisc rotation law given in \citet{mah03}. Also, because $|B_{\phi, D}| < |B_{\phi, W}|$, the strength of $B_{\phi}$ decreases slightly across the shock boundary.

\section{Equations for Protodisc Regions}\label{sec-disc-proto-eqns}
Here, we set up equations for the structure of protodiscs and obtain analytic solutions. We do not consider a disc region to be a protodisc after the onset of MRI. Since material flows continuously into a protodisc region along magnetic flux tubes and does not flow outwards, its height $h$ increases with time $t$. The accumulation of mass can be expressed in the form
\begin{equation}\label{eqn-disc-proto-mass-accum}
\int_0^h \, \rho \, dz \, = \, \rho_W v_{n, W} \, t \, .
\end{equation}

Let $X$ be a point on the equatorial plane at distance $\varpi$ from the centre of the star and let $Q$ be on the shock surface vertically above $X$. To obtain values of the magnetic field components at the shock boundary, we assume that during the protodisc phase $h$ is relatively small compared with $\varpi$ and that the angle $\delta$ is small. The computational results that we obtain for our models confirm that this assumption is appropriate. Then, we have $B_{\varpi, D} = B_{n, D} \sin\delta \approx 0$ and $B_{z, D} = -B_{n, D} \cos\delta \approx -B_{m, W}$. In such protodisc regions, the meridional magnetic field, which is a continuation of the dipole-type field in the wind zone, is almost vertical and we may take $B_\varpi \approx 0$ and $B_z \approx -B_{m, W}$.

The general form of the azimuthal component of the momentum equation for an axially symmetric disc region is
\begin{equation}\label{eqn-disc-proto-mom-phi-0}
\frac{\partial u_\phi}{\partial t} = -u_\varpi \frac{\partial
u_\phi}{\partial \varpi} - u_z \frac{\partial u_\phi}{\partial z}
- \frac{u_\varpi u_\phi}{\varpi} + \frac{\mathcal{F}_\phi}{\rho} +
\frac{1}{\rho \varpi^2} \,
\frac{\partial}{\partial\varpi}\left(\varpi^2 \alpha \, \rho \,
c_s^2 \right)\, ,
\end{equation}
where $\alpha$ is the viscosity coefficient developed by
\citet{sha73}, $c_s$ is the local sound speed and
\begin{equation}\label{eqn-magnetic-force-phi-comp-0}
\mathcal{F}_\phi = \frac{1}{4 \pi \varpi} \left[B_\varpi \, \frac{\partial}{\partial \varpi}\left(\varpi B_\phi \right) + B_z \, \frac{\partial}{\partial z}\left( \varpi B_\phi\right)\right]
\end{equation}
is the $\phi$-component of the magnetic force. Before the onset of MRI, material in a protodisc region and the magnetic field rotate steadily with super-Keplerian speed and $\alpha =0$. Since $\bvect{u} = (0, u_\phi, 0)$, we require solutions that satisfy the conditions $\partial \bvect{u_\phi}/\partial t = 0$, $\partial \bvect{B}/\partial t = \bvect{0}$ and $\textrm{div} \bvect{B} = 0$.  Then, equation (\ref{eqn-disc-proto-mom-phi-0}) requires that $\partial B_\phi/\partial z = 0$, which says that the azimuthal component of the magnetic field between the equatorial plane and the upper shock boundary is constant along a vertical line. Applying the boundary condition at
the shock surface we have $B_\phi = B_{\phi, D}$. Note that, in the part of the protodisc below the equatorial plane, $B_\phi$ will be positive and have the same magnitude as in the upper part where it is negative. This implies a discontinuity in $B_\phi$ across the equatorial plane, which is physically plausible and is accounted for by the presence of a current sheet. The condition that $\partial \bvect{B}/\partial t = \bvect{0}$ together with the MHD induction equation imply that $\textrm{curl}(\bvect{u \times B)} = \bvect{0}$. When we substitute the solutions that we have already obtained for $B_\varpi$, $B_\phi$ and $B_z$ we find that $u_\phi$ satisfies the condition $\partial u_\phi/\partial z = 0$. Therefore, as in the case of $B_\phi$, the rotation speed $u_\phi$ is the same at all points along a vertical line in the protodisc and using the boundary condition at the shock surface, we obtain $u_\phi = u_{\phi, D}$. Thus, we have a physically valid model of
a protodisc that is described by the self-consistent solutions $\bvect{u} = (0, u_{\phi, D}, 0)$ and $\bvect{B} = (0, B_{\phi, D}, -B_{m, W})$.  Also, we find that the vertical component of the magnetic force $\mathcal{F}_z = 0$ in the protodisc region of our model and the radial component $\mathcal{F}_{\varpi, X} \approx {(b-1) \, B_{m, W}^2}/{4 \pi \, \varpi}$ at the point $X$ on the equatorial plane. Since $u_z$ is negligible in the protodisc region, the general form of the $z$-component of the momentum equation for such a region is
\begin{equation}\label{eqn-disc-proto-mom-z-0}
\frac{\partial p}{\partial z} = - \frac{\rho \, GM}{r^2} \, \frac{z}{r} \, - \frac{1}{4\pi} \, \left[\left(\frac{\partial B_\varpi}{\partial z} - \frac{\partial B_z}{\partial
\varpi}\right)B_\varpi + \left(\frac{\partial B_\phi}{\partial z}\right) B_\phi \right]\, ,
\end{equation}
where $p$ and $\rho$ are functions of time $t$. This equation implies that the pressure gradient in the $z$-direction at any point is instantaneously balanced by the $z$-components of stellar gravity and magnetic force. In the case of our protodisc models in which $z$ is much smaller than $r$ we may take $z/r^3 \approx z/\varpi^3$. Since $p = \rho \, a^2$ and $a$ is independent of $z$, equation (\ref{eqn-disc-proto-mom-z-0}) gives
\begin{equation}\label{eqn-disc-proto-mom-z-2}
\frac{\partial \rho}{\partial z} = - \frac{2z}{H^2} \, \rho \, .
\end{equation}
where $H = H(\varpi) = \left(2 \varpi^3 a^2 \, / GM\right)^{1/2}$ is the density scale height in the protodisc. This equation implies that the density gradient of the protodisc in the vertical direction is influenced by the vertical component of gravity. The vertical component of the magnetic force is zero in our model and does not contribute to vertical equilibrium. We solve this equation to write the density of the protodisc at height $z$ above $X$, in the form
\begin{equation}\label{eqn-disc-proto-rho-Z} \rho = \rho_D \, e^{\left(h^2-z^2\right)/H^2} \, ,
\end{equation}
where $\rho_D$ is the density of the protodisc at the point $Q$ of the shock boundary. Using equation (\ref{eqn-disc-proto-mass-accum}) and the expression for $\rho$ in
equation (\ref{eqn-disc-proto-rho-Z}) we obtain an equation connecting the height $h$ of the protodisc and time $t$ in the form
\begin{equation}\label{eqn-disc-proto-height-time-t}
t = \frac{H}{\rho_W v_{n, W}}\left[\frac{\sqrt{\pi}}{2} {\rho_D} \, e^{h^2/H^2} \, \erf\left(\frac{h}{H}\right) \right] \, ,
\end{equation}
where $\erf(h/H)$ denotes the error function value at $h/H$.

Equations (\ref{eqn-disc-proto-rho-Z}) and (\ref{eqn-disc-proto-height-time-t}) show that the density at any point $X$ on the equatorial plane of the protodisc region increases with time. Then, unless MRI intervenes and terminates the protodisc phase, the density at $X$ will increase to the value $\rho_{X, edge}$ that is required for the centrifugal force to
balance the sum of gravity, magnetic force and thermal pressure. That is, $\rho_{X, edge}$ is obtained as the solution of the equation
\begin{equation}\label{eqn-disc-proto-mom-radial-comp-0}
\rho_{X, edge} \left( \frac{{u}_{\phi}^2}{\varpi} -
\frac{GM}{\varpi^2}\right) - \frac{\partial \left(\rho
a^2\right)}{\partial\varpi } - \frac{(b-1) \, B_{m, W}^2}{4 \pi \,
\varpi} = 0 \, .
\end{equation}
We denote the time at which this occurs by $t_{X, edge}$ and let $h_{X, edge}$ be the height of the protodisc at $X_{edge}$ at this time. The value of $h_{X, edge}$ is obtained by putting $\rho = \rho_{X, edge}$ in equation (\ref{eqn-disc-proto-rho-Z}) and
solving for $h$. The time $t_{X, edge}$ at which this occurs is found by using equation (\ref{eqn-disc-proto-height-time-t}). In addition to giving us the values of the properties at the outer edge, the importance of these equations is that they give the rate at which the radius $x_{edge}$ of the outer edge of the protodisc decreases. Thus, we are able to compute the radial extent of a protodisc at the time at which MRI occurs. Numerical values are shown in the tables presented in section \ref{sec-disc-proto-applics-results}.

In our model of a protodisc, a magnetic field with a sufficiently strong meridional component channels the flow of wind material into an equatorial disc region. A much weaker azimuthal component of the magnetic field acts in concert with the meridional component to apply a torque on the wind material so that it enters the protodisc region with super-Keplerian angular velocity after crossing its shock boundary. The radial component of the magnetic force in the protodisc is initially able to assist gravity in
maintaining radial equilibrium against the centrifugal force and thermal pressure. The protodisc region rotates steadily with no meridional motion when $u_\phi$ and $B_\phi$ are constant along vertical lines. Consequently, the vertical and azimuthal components of the magnetic force are zero in the protodisc region and the density gradient in the vertical direction is influenced only by stellar gravity. Increase in equatorial density causes centrifugal outflow at the outer edge and causes the radial extent of the protodisc to decrease. It will continue to do so unless MRI or some other MHD instability occurs. We discuss this in section \ref{sec-onset-mri-protodiscs}.

\section{Onset of MRI in Protodiscs}\label{sec-onset-mri-protodiscs}
As we pointed out in sections \ref{sec-proto-region} and \ref{sec-shock-proto-jump-conditions}, continuous flow of wind material causes continuous shock heating of the protodisc. The inflow time of the wind into the protodisc region is
\begin{equation}\label{eqn-disc-proto-inflow-time}
t_{inflow} \approx \frac{h}{v_n + U_s} \, .
\end{equation}
The temperature of a protodisc will remain high if the shock heating is not offset by radiative cooling. The radiative cooling time for the disc material is given by $t_{cool}=
{\mathcal{U}}/{{L_{cool} }} = {{3kT}}/({2N_e \, \Lambda })$ s, where $\mathcal{U} = 3kTN_{total}/2$ $\textrm{ergs/cm}^3$ is the internal energy per unit volume and $L_{cool}  = N_e \, N_{total} \, \Lambda$ $\textrm{ergs}/(\textrm{cm}^3 \textrm{s})$ is the cooling rate per unit volume, where $N$ denotes the number of particles per cubic centimetre. \citet{cox00} gives $\log \Lambda =  - 21.65 - 0.6\left(\log T - 5.5\right)$. From this we obtain the expression $\Lambda  = \left({10^{18.35} \, T^{0.6}}\right)^{-1}$  ergs $\textrm{cm}^3$/s for the cooling function that is appropriate for the postshock temperature ranges of our protodiscs. We assume that the disc material consists of Hydrogen and Helium with mass fractions $X_H$ and $1 - X_H$, respectively. Then the number density of electrons is given by $N_e  = (1 + X_H) \rho _D /(2 m_H)$ per cubic centimetre and the
cooling time is
\begin{equation}\label{eqn-disc-cool-time} t_{cool} \, = \, \frac{{6.72\; \times 10^{18} \;k\;m_H \; T_D^{1.6}}}{{(1 + X_H)\;\rho _D \;}}\; = \;\frac{{1.55 \times 10^{- 21} \;\;T_D^{1.6} }}{{(1+X_H) \rho _D \;}} \quad \textrm{s}.
\end{equation}
Our computations show that the radiative cooling rate in our models is slower than the rate of shock heating so that protodiscs continue to maintain their high temperatures.

There are several values of the density and time scales that play critical roles in the evolution of protodiscs. The requirement $\beta_{mag} > 1$ for the onset of MRI stipulated in condition (\ref{eqn-disc-proto-MRI-conditions}) will be satisfied when the density $\rho_X$ at a point $X$ on the equatorial plane exceeds the critical value $\rho_{X,MRI}$ given by
\begin{equation}\label{eqn-disc-proto-MRI-density}
\rho_{X, MRI} = \frac{B_X^2}{8 \pi a^2} = \frac{B_{m, W}^2}{8 \pi a^2} \, ,
\end{equation}
because $B_{\phi, X}=0$. If $\rho_{X, MRI} < \rho_{X, edge}$ at a point $X$ in a protodisc, the value of $\rho_X$  will reach the value of the density required for MRI to set in at $X$ before the outer edge of the protodisc reaches the point $X$.  Let $t_{X, MRI}$ denote the time required for MRI to set in at the point $X$.

Let the point $P$ on the stellar surface be linked to the point $X$ on the equatorial plane of a protodisc region by a meridional magnetic field line. Let $\omega$ be the angular velocity of the stellar surface at $P$ and let $\Omega$ be that of the protodisc at $X$. In section \ref{sec-proto-region}, we pointed out that in models of protodiscs where the shock surface is stationary, the condition that $\omega$ should increase towards the equator along the stellar surface will ensure that the second necessary condition, ${\partial \, \Omega}/{\partial\varpi} < 0$, for the onset of MRI is satisfied. However, in our models of protodiscs, the shock surface moves so that its height increases with time. In this case, equations (\ref{eqn-wind-v-phi-omega}), (\ref{eqn-shock-disc-proto-Bn}) and
(\ref{eqn-shock-disc-proto-uphiD}) give
\begin{equation}\label{eqn-proto-relation-Omega-omega}
\Omega = \omega -  \frac{v_{n, W} \, B_{\phi, W} \, U_s^2}{\varpi \, B_{n, W}\left(\mathcal{A}_{n, D}^2 - U_s^2\right)} \, ,
\end{equation}
where we have used the equations $\Omega \, \varpi = u_\phi = u_{\phi, D}$. We interpret this to be a generalised form of the protodisc rotation law. Note that we recover the protodisc rotation theorem in \citet{mah03} when $U_s = 0$. For stars with differential rotation, we need the distribution of $\omega$ along the stellar surface to compute values of $\Omega$. To determine whether $\Omega$ will decrease with increasing $\varpi$ in a
protodisc for realistic values of $\omega$, we use the angular velocity model for the Sun (e.g., see Solar Rotation in Allen's Astrophysical Quantities 2000), which has the form
\begin{equation}\label{eqn-proto-star-omega-model}
\omega = \omega_{eq}\left({1 - \epsilon \cos^2\theta_\star}\right) = \omega_{eq}\left[1 - \epsilon + \frac{\epsilon}{x^{(b-2)}}\right] \, ,
\end{equation}
where $\omega_{eq}$ is the angular velocity of the star at the equator, $\theta_\star$ denotes colatitude of the point $P$ on the stellar surface, $x$ is the non-dimensional distance of the point $X$ from the rotation axis and $\epsilon$ is a constant such that $0 \leq \epsilon < 1$. We have used the relation $\sin^2 \theta_\star = 1/x^{(b-2)}$ for dipole-type fields \citep{mah03}. When $\omega$ varies along the stellar surface, so does the rotation rate $\mathcal{S}_\star$. We use $\mathcal{S}_{\star, eq}$ to denote the value of
$\mathcal{S}_\star$ corresponding to $\omega_{eq}$. In section \ref{sec-disc-proto-applics-results}, we present results of numerical computations for a variety of models. We find that the condition ${\partial \, \Omega}/{\partial\varpi} < 0$ is satisfied in protodisc regions for a wide range of permissible values of $\epsilon$.

\begin{table*}
\begin{minipage}{170mm}
\caption{Properties of the different stellar models considered}
\label{table-star-properties}  \begin{tabular}{llrcccrccc}  \hline
{Star} & {Spectral} & {$M$} & {$R$} & {$T_{eff}$}
 & {${\dot M}$} & {$V_\infty$} & {$\rho_\circ$}
& {$v_{m,\circ}$} & {${v_{\phi,\star,eq,crit}}$} \\
 {Name} & {Type} & {($M_\odot$)}
& {($R_\odot$)} & {($10^4$K)} & {($10^{-9} M_\odot
\textrm{yr}^{-1}$)} & {(${{\textrm {km s}}^{-1}}$)} &
{($\textrm{gm cm}^{-3}$)} & {($10^6\, \textrm{cm s}^{-1}$)} &
{($\textrm{km s}^{-1}$)} \\\hline
Model O6.5 &O6.5 & $29\;\;$  & $10\;\;$  & $4.0$  &  $310$    & $2500$ &$1.38 \times 10^{-12}$ & $2.32$ & $661$  \\
Model B0 & B0   & $15\;\;$  & $6.6$    & $3.2$  &   $27$    & $1300$ & $3.09 \times 10^{-13}$ & $2.08$ & $634$  \\
Model B2 & B2   & $8.3$     & $4.5$    & $2.3$  &  $0.4$    & $ 840$ & $1.16 \times 10^{-14}$ & $1.76$ & $589$  \\
Model B5 & B5   & $4.5$     & $3.5$    & $1.5$  &  $0.01$   & $ 580$ & $5.95 \times 10^{-16}$ & $1.42$ & $495$ \\
Model B9 & B9   & $2.6$     & $2.6$    & $1.0$  &  $0.0013$ & $ 460$ & $1.72 \times 10^{-16}$ & $1.16$ & $437$ \\[4pt]
$\alpha$ Ara & B2   & $9.8\;$  & $4.8$    & $2.3$  &   $0.45$    & $1540$ & $1.15 \times 10^{-14}$ & $1.76$ & $625$  \\
$\alpha$ Eri & B3   & $6.9\;$  & $4.1$    & $1.9$  &   $0.042$    & $1330$ & $1.63 \times 10^{-15}$ & $1.62$ & $569$  \\
$\alpha$ Col & B7   & $3.7\;$  & $3.4$    & $1.3$  &   $0.003$    & $1250$ & $2.06 \times 10^{-16}$ & $1.32$ & $457$  \\
\hline
\end{tabular}

Entries for the stellar properties are from \citet{bjo93} for the theoretical models and from \citet{coh97} for specific stars. Here, ${v_{\phi,\star,eq,crit}}$ is the critical Keplerian rotation speed at the equator.
\end{minipage}
\end{table*}

\section{Results for Protodisc Models}\label{sec-disc-proto-applics-results}
We have applied the structure equations and shock jump conditions to study protodiscs around a variety of stellar models with properties listed in Table \ref{table-star-properties}. We have performed computations for uniform and differential rotation with different stellar rotation rates and different surface magnetic field strengths. We have used equation (\ref{eqn-proto-relation-Omega-omega}) to compute values of the angular velocity $\Omega$ of protodiscs around stellar models with solar-type differential rotation given by equation (\ref{eqn-proto-star-omega-model}), for different values of $\epsilon$. In these models, the stellar angular velocity $\omega$ increases outwards along the surface towards the equator. We display numerical results for protodiscs around stellar models with uniform rotation and for solar-type differential rotation with $\epsilon = 0.2$. In section \ref{sec-windzone} we mentioned that $b$ will be less than 3 in the wind zone. Here, we show results for $b = 3$ because they are not significantly different from those for models with values of $b$ slightly less than 3.
Table \ref{table-protodisc-properties-alpha-ara-differential-Omega} shows the values of the angular velocity $\Omega$ of protodiscs around differentially rotating stellar models with the parameters of $\alpha$ Ara, dipole $b=3$, $\beta_{vel}=1$, $\mathcal{S}_{\star, eq} = 0.7$ and $\omega_{eq} = 1.312 \times 10^{-4}$ radians per second. These results show that
$\partial\Omega/\partial\varpi < 0$ for values of $x$ between $x_{inner}$ and $x_2$. We find that in models with $\omega$ given by equation (\ref{eqn-proto-relation-Omega-omega}) and $\epsilon \ga 0.1$, the condition $\partial \Omega/\partial \varpi < 0$ for the onset of MRI is satisfied. In the case of a stellar model with uniform rotation where $\omega$ is constant along the surface, our numerical results show that the value of the angular velocity $\Omega$ of the protodisc is constant to an accuracy of $0.1$\% or better for B-type stars and for O-type stars the variation in the value of $\Omega$ is of the order of a few percent.

\begin{table*}
\begin{minipage}{120mm}
\caption{Variation of angular velocity $\Omega$ and values of $x_2$ for protodiscs of differentially rotating stellar models with the parameters of $\alpha$ Ara, dipole $b=3$,
$\beta_{vel}=1$,  $\mathcal{S}_{\star, eq} = 0.7$, $\omega_{eq} = 1.312 \times 10^{-4}$ radians per second and $\epsilon = 0.2$}
\label{table-protodisc-properties-alpha-ara-differential-Omega}
 \begin{tabular}{lcllcllc}  \hline \multicolumn{2}{c}{$B_{m, \circ} = 57G$, \, $x_2 =
2.31$} & {} &\multicolumn{2}{c}{$B_{m, \circ}=119G$, \, $x_2 =
2.37$} & {} &\multicolumn{2}{c}{$B_{m, \circ}=245G$, \, $x_2 =
2.45$}
\\
\cline{1-2} \cline{4-5} \cline{7-8}
\\
 {$x$} & {$\Omega$} & {} & {$x$} & {$\Omega$} & {} & {$x$} & {$\Omega$}\\
 {} & {radians/s} & {} & {}
& {radians/s} & {} &  {} & {radians/s}\\ \hline
$1.31$  & $ 1.250 \times 10^{-4}$  &  &$1.31$  & $ 1.250 \times 10^{-4}$  & & $1.31$  & $1.250 \times 10^{-4}$\\
$1.61$  & $ 1.213 \times 10^{-4}$  &  &$1.61$  & $ 1.213 \times 10^{-4}$  & & $1.61$  & $1.213 \times 10^{-4}$\\
$2.01$  & $ 1.195 \times 10^{-4}$  &  &$2.01$  & $ 1.183 \times 10^{-4}$  & & $2.01$  & $1.181 \times 10^{-4}$\\
$2.31$  & $ 1.192 \times 10^{-4}$  &  &$2.31$  & $ 1.168 \times 10^{-4}$  & & $2.41$  & $1.159 \times 10^{-4}$\\
$2.41$  & $ 1.194 \times 10^{-4}$  &  &$2.41$  & $ 1.165 \times 10^{-4}$ & & $2.51$  & $1.156 \times 10^{-4} $ \\
\hline
\end{tabular}
\end{minipage}
\end{table*}

In all our models, the outer edge $X_{edge}$ of the protodisc initially coincides with the point $X_1$ where the sum of the magnetic force and gravity is equal to the centrifugal force. The rotation speed at points along a protodisc is super-Keplerian before the onset of MRI. As material continues to flow from the wind zone into the protodisc region, the density along the equatorial plane increases and material at the outer edge starts
to flow away radially so that $X_{edge}$ moves towards the inner boundary $X_{inner}$. In models with uniform rotation, the protodisc will continue to shrink radially unless some MHD instability intervenes. The rate at which $X_{edge}$ approaches $X_{inner}$ becomes slower as it gets closer to $X_{inner}$. In some of the cases the time taken for $X_{edge}$ to approach $X_{inner}$ can be many years. We use $t_{shrink}$ to denote the value of $t_{edge}$ when a protodisc is not affected by any instability and shrinks to a thin ring. In such a case, the radial extent of the protodisc will be small during most of this time and the protodisc will have the appearance of a ring. Our numerical results confirm that the values of $h$ are small compared with $\varpi$ during the protodisc phase in all the models that we consider.

In a protodisc where the angular velocity $\omega$ of the central star increases outwards along the stellar surface towards the equator, we find that as $X_{edge}$ moves inwards, it reaches a point $X_2$ at the same time $t_2$ at which MRI sets in at $X_2$. At all points between $X_{inner}$ and $X_2$ we find that MRI sets in before the local density increases to the value required for centrifugal outflow. We do not follow the evolution of disc
material after the onset of MRI when viscosity becomes important. In some cases where MRI sets in after the protodisc has shrunk considerably, a quasi-steady ring may be formed.

\begin{table*}
\begin{minipage}{158mm}
 \caption{Properties of protodiscs of theoretical stellar models listed in Table \ref{table-star-properties} with uniform rotation or differential rotation and aligned dipole-type magnetic fields}
\label{table-protodisc-properties-bc-cohen}
 \begin{tabular}{lcrrrrrrrrrrrr}  \hline \multicolumn{4}{c}{} &\multicolumn{4}{c}{Uniform
Rotation with $\mathcal{S}_\star = \mathcal{S}_{\star, eq}$} & {}
&\multicolumn{5}{c}{Differential Rotation with $\epsilon = 0.2$}
\\ \cline{5-8} \cline{10-14}
\\  {Star}
& {$\mathcal{S}_{\star, eq}$} & {$B_{m,\circ}$} & {$x_{lim}$} &
{$B_{\phi,\circ}$} & {$x_{inner}$} & {$x_1$} & {$t_{shrink}$} & {}
& {$B_{\phi,\circ}$} & {$x_{inner}$} & {$x_1$} & {$x_2$} & {$t_2$}
\\
 {Model} & {} & {(G)} & {}
& {(G)} & {} & {} & {(days)} & {} & {(G)} & {} & {} & {} &
{(days)}\\ \hline
O6.5 & $0.50$   & $ 194$  & $ 1.78$  & $ -26.81$  & $ 1.59$  & $ 1.74$   & $ 0.03$   & &$ -24.47$  & $ 1.68$  & $ 1.77$  & $ 1.76$  & $   0.006$\\
  &   & $ 432$  & $ 2.61$  & $ -36.09$  & $ 1.59$  & $ 2.36$ & $   0.22$  & &$ -31.64$  & $ 1.68$  & $ 2.43$  & $ 2.42$  & $   0.007$\\
  & $0.70$  & $ 194$  & $ 1.78$  & $ -37.54$  & $ 1.27$  & $ 1.58$ & $   0.18$   & &$ -34.25$  & $ 1.31$  & $ 1.62$  & $ 1.61$  & $   0.006$ \\
  &         & $ 432$  & $ 2.61$  & $ -50.52$  & $ 1.27$  & $ 2.14$ & $   0.77$  & &$ -44.29$  & $ 1.31$  & $ 2.22$  & $ 2.21$  & $
  0.007$\\[2pt]
B0    & $0.50$  & $ 104$  & $ 2.25$  & $ -17.72$  & $ 1.59$  & $ 1.92$ & $ 0.15$  & &$ -15.74$  & $ 1.68$  & $ 2.01$  & $ 2.00$  & $   0.006$\\
         &        & $ 225$  & $ 3.39$  & $ -24.32$  & $ 1.59$  & $ 2.50$  & $ 0.71$   & &$ -20.89$  & $ 1.68$  & $ 2.67$  & $ 2.66$  & $   0.010$\\
         &        & $ 473$  & $ 5.23$  & $ -34.83$  & $ 1.59$  & $ 3.22$  & $ 2.88$   & &$ -29.20$  & $ 1.68$  & $ 3.65$  & $ 3.12$  & $   0.093$\\
         & $0.70$ & $ 225$  & $ 3.39$  & $ -34.05$  & $ 1.27$  & $ 1.60$  & $ 0.39$  & &$ -29.25$  & $ 1.31$  & $ 2.08$  & $ 1.51$  & $   0.334$\\
         &        & $ 473$  & $ 5.23$  & $ -48.76$  & $ 1.27$  & $ 2.22$  & $ 7.38$  & &$ -40.88$  & $ 1.31$  & $ 2.62$  & $ 1.57$  & $   1.039$\\[{2pt}]
B2 & $0.50$ & $  58$  & $ 4.79$  & $  -6.47$  & $ 1.59$  & $ 2.14$  & $ 2.07$  & &$  -5.44$  & $ 1.68$  & $ 2.53$  & $ 2.10$  & $   0.271$\\
         &        & $ 121$  & $ 7.46$  & $  -9.54$  & $ 1.59$  & $ 2.64$  & $ 8.02$  & &$  -7.88$  & $ 1.68$  & $ 3.07$  & $ 2.16$  & $   0.969$ \\
         &        & $ 247$  & $11.70$  & $ -14.44$  & $ 1.59$  & $ 3.16$  & $ 31.18$  & &$ -11.80$  & $ 1.68$  & $ 3.65$  & $ 2.20$  & $   3.542$ \\
         & $0.70$ & $ 121$  & $ 7.46$  & $ -13.35$  & $ 1.27$  & $ 1.70$ & $ 16.22$  & &$ -11.04$  & $ 1.31$   & $ 2.00$  & $ 1.37$  & $   7.580$ \\
         &        & $ 247$  & $11.70$  & $ -20.22$  & $ 1.27$  & $ 2.26$  & $ 63.36$  & &$ -16.52$ & $ 1.31$  & $ 2.60$  & $ 1.37$  & $  29.788$ \\[{2pt}]
B5 & $0.50$ & $  62$  & $15.13$  & $  -3.76$  & $ 1.59$  & $ 2.94$ & $ 68.00$  & &$  -3.06$  & $ 1.68$  & $ 3.39$  & $ 1.96$  & $  12.335$\\
         &        & $ 125$  & $23.89$  & $  -5.84$  & $ 1.59$  & $ 3.54$  & $ 266.14$  & &$  -4.72$ & $ 1.68$  & $ 4.07$  & $ 2.00$  & $  42.823$ \\
         &        & $ 253$  & $37.80$  & $  -9.13$  & $ 1.59$  & $ 4.24$  & $ 1049.25$  & &$  -7.35$  & $ 1.68$  & $ 4.85$  & $ 2.00$  & $ 170.223$ \\
         & $0.70$ & $  62$  & $15.13$  & $  -5.27$  & $ 1.27$  & $ 2.10$ & $ 119.78$  & &$  -4.28$  & $ 1.31$  & $ 2.40$  & $ 1.35$  & $  74.824$\\
         &        & $ 125$  & $23.89$  & $  -8.17$  & $ 1.27$  & $ 2.64$ & $ 474.30$  & &$  -6.60$  & $ 1.31$  & $ 3.00$  & $ 1.35$  & $ 296.062$ \\
         &        & $ 253$  & $37.80$  & $ -12.79$  & $ 1.27$  & $ 3.22$ & $1885.03$  & &$ -10.30$  & $ 1.31$  & $ 3.66$  & $ 1.35$  & $1175.860$ \\[{2pt}]
B9 & $0.50$ & $  63$  & $26.30$  & $  -3.08$  & $ 1.59$  & $ 3.36$ & $ 290.95$  & &$  -2.49$  & $ 1.68$  & $ 3.85$  & $ 1.90$  & $  61.286$\\
         &        & $ 126$  & $41.63$  & $  -4.83$  & $ 1.59$  & $ 4.04$  & $ 1146.13$  & &$  -3.88$  & $ 1.68$  & $ 4.61$  & $ 1.92$  & $ 224.1299$ \\
         &        & $ 254$  & $65.96$  & $  -7.60$  & $ 1.59$  & $ 4.80$  & $ 4540.98$ & &$  -6.10$  & $ 1.68$  & $ 5.49$  & $ 1.92$  & $ 893.279$ \\
         & $0.70$ & $  63$  & $26.30$  & $  -4.31$  & $ 1.27$  & $ 2.48$ & $ 479.826$  & &$  -3.48$  & $ 1.31$  & $ 2.84$  & $ 1.33$  & $ 4867.086$ \\
         &        & $ 126$  & $41.63$  & $  -6.76$  & $ 1.27$  & $ 3.06$ & $1905.91$  & &$  -5.44$  & $ 1.31$  & $ 3.48$  & $ 1.35$  & $1177.084$ \\[2pt]
$\alpha$ Ara & $0.50$  & $  57$ & $ 3.97$  & $  -4.29$  & $ 1.59$   & $ 3.026$  & $0.67$  & &$ -3.64$   & $ 1.68$  & $ 3.21$  & $ 3.20$  & $   0.004$\\
         &   & $ 119$  & $ 6.15$  & $  -6.22$  & $ 1.59$  & $ 4.16$  & $ 2.68$ & &$  -5.18$  & $ 1.68$  & $ 4.45$  & $ 4.26$  & $   0.026$\\
         &   & $ 245$  & $ 9.63$  &$  -9.33$  & $ 1.59$  & $ 4.92$  & $ 10.61$ & &$  -7.66$  & $ 1.68$  & $ 5.81$  & $ 4.30$  & $   0.202$\\
   & $0.70$  & $  57$  & $ 3.97$  &$  -6.00$  & $ 1.27$  & $ 2.64$  & $ 1.96$   & &$  -5.10$  & $ 1.31$ & $ 2.82$  & $ 2.31$  & $   0.048$\\
         &   & $ 119$  & $ 6.15$  &$  -8.71$  & $ 1.27$  & $ 3.08$  & $ 7.63$ & &$  -7.26$  & $ 1.31$   & $ 3.64$  & $ 2.37$  & $   0.209$\\
         &   & $ 245$  & $ 9.63$  &$ -13.06$  & $ 1.27$  & $ 3.60$  & $ 29.84$ & &$ -10.72$  & $ 1.31$  & $ 4.20$  & $ 2.45$  & $   0.765$\\[{2pt}]
$\alpha$ Eri & $0.50$  & $  60$  & $ 7.97$  &$  -2.77$  & $ 1.59$  & $ 4.44$  & $ 5.88$  & &$  -2.28$   & $ 1.68$  & $ 5.21$  & $ 3.98$  & $   0.126$\\
        &    & $ 124$  & $12.52$  &$  -4.20$  & $ 1.59$  & $ 5.06$  & $ 23.22$ & &$  -3.43$  & $ 1.68$  & $ 5.93$  & $ 4.04$  & $   0.535$\\
        &    & $ 251$  & $19.75$  &$  -6.49$  & $ 1.59$  & $ 5.84$  & $ 91.96$ & &$  -5.26$  & $ 1.68$  & $ 6.79$  & $ 4.12$  & $   2.044$\\
   & $0.70$  & $  60$  & $ 7.97$  &$  -3.87$  & $ 1.27$  & $ 3.20$  & $ 16.24$ & &$  -3.19$  & $ 1.31$  & $ 3.74$  & $ 2.17$  & $   0.604$\\
        &    & $ 124$  & $12.52$  &$  -5.88$  & $ 1.27$  & $ 3.78$  & $ 63.68$ & &$  -4.80$  & $ 1.31$  & $ 4.36$  & $ 2.25$  & $   2.122$\\
        &    & $ 251$  & $19.75$  &$  -9.08$  & $ 1.27$  & $ 4.44$  & $ 251.31$ & &$  -7.36$ & $ 1.31$  & $ 5.12$  & $ 2.33$  & $   7.462$\\[{2pt}]
$\alpha$ Col & $0.50$  & $  62$  & $17.10$  &$  -1.50$  & $ 1.59$  & $ 6.36$  & $ 54.50$ & &$  -1.22$  & $ 1.68$ & $ 7.45$  & $ 5.02$  & $   0.807$\\
        &    & $ 126$  & $27.02$  &$  -2.33$  & $ 1.59$  & $ 7.32$ & $ 216.70$ & &$  -1.88$  & $ 1.68$  & $ 8.51$  & $ 5.10$  & $   3.104$\\
        &    & $ 253$  & $42.76$  &$  -3.66$  & $ 1.59$  & $ 8.46$ & $ 863.29$  & &$  -2.94$  & $ 1.68$   & $ 9.83$  & $ 5.18$  & $  11.809$\\
        & $0.70$  & $  62$  & $17.10$  &$  -2.10$  & $ 1.27$  & $ 4.80$  & $ 158.30$ & &$  -1.70$  & $ 1.31$  & $ 5.58$  & $ 3.05$  & $ 2.243$\\
        &    & $ 126$  & $27.02$  &$  -3.27$  & $ 1.27$  & $ 5.62$ & $ 627.78$ & &$  -2.64$  & $ 1.31$   & $ 6.50$  & $ 3.11$  & $   8.402$\\
        &    & $ 253$  & $42.76$  &$  -5.12$  & $ 1.27$  & $ 6.56$ & $ 2497.14$ & &$  -4.12$  & $ 1.31$    & $ 7.58$  & $ 3.17$  & $  31.504$\\
\hline
 \end{tabular}

Stellar models refer to those listed in Table \ref{table-star-properties}.  In these models, dipole $b=3$ and wind $\beta_{vel}=1$. In all these models, $T_D$ is of order $1 \,  \textrm{to} \, 10\times 10^6$K, \,  $U_s$ is in the range from 50 to 200 kms, $t_{inflow} \sim 1 \, \textrm{to} \, 40$ mins. The values of $t_{cool}$ are of the order 10 hours to a few days and depend on the properties of the stars considered. In the case of the last three stellar models, they are of the order of a few days to several months.
\end{minipage}
\end{table*}

Table \ref{table-protodisc-properties-bc-cohen} shows the radial extents of protodiscs and the locations and times at which MRI occurs in our protodisc models. The values of $x_{inner}$, $x_{lim}$, $x_{1}$ and $x_{2}$ and the time $t_{2}$ are shown for protodiscs of different stellar models with uniform ${\mathcal S}_\star$ equal to $0.5$ and $0.7$, and for differential rotation with ${\mathcal S}_{\star, eq} = 0.7$ and $\epsilon = 0.2$. In
most cases, MRI sets in before the protodisc shrinks to a thin ring and occurs on a time scale of hours or days. In situations where the radial extents of protodiscs decrease to small values, we find that the time taken may vary from several days to a few years depending on the type of star, its field strength and its rotation rate. In all our models, the protodisc temperature $T_D$ is of the order of $1 \, \textrm{to} \, 10 \times 10^6$K and the wind inflow time $t_{inflow}$ has values in the range from about 1 minute to about 40 minutes. We use equation (\ref{eqn-disc-cool-time}) to derive the expression $t_{cool} = 1.82\times10^{-18} T^{1.6}/\rho$ seconds for the radiative cooling time for a plasma with $70\%$ Hydrogen and $30\%$ Helium by mass. In all our models, we find that the values of $t_{cool}$ have values in a range from about 10 hours to several days. Thus, $t_{cool}$ is many orders of magnitude larger than the inflow time $t_{inflow}$ of wind material. Also, in all our protodisc models, the values of $t_{cool}$ are much longer than $t_{edge}$ when $x$ is large. Thus, the outer parts of protodiscs do not cool down before the material in those parts flow away from the protodisc. The speed $U_{shock}$ of the shock boundary is of the order of 10 to 100 kilometres per second and depends on the stellar model and the radial distance from the axis rotation. In all the protodiscs, the values of $U_{shock}$ are approximately $30\%$ of the values of the wind flow speed $v_{n, W}$. An important feature of the results is that for given values of the rotation speed and surface magnetic field strengths, when the wind speed is faster the values of $x_2$ are larger. This is because wind flows that are faster lead to greater heating of protodiscs causing earlier onset of MRI. This is apparent when we compare the protodisc properties for the theoretical model B2 star and the models corresponding to $\alpha$ Ara  given in table \ref{table-protodisc-properties-bc-cohen}.

In stellar models of earlier types such as O6.5 and B0 shown in table \ref{table-protodisc-properties-bc-cohen}, we find that MRI occurs almost instantaneously so that the protodisc phase of the disc region ends almost as soon as it starts. In the case of later type stars such as B5 or B9 shown in the same table, the time taken for
MRI to occur can be several months or years depending on the field strength. For most of this time, the radial extents of such protodiscs will be small because the rate at which they shrink is faster initially and it slows down drastically as $X_{edge}$ approaches $X_{inner}$. Note that there is a difference in the time taken for the onset of MRI between the protodiscs of these models and the model corresponding to the B7 star $\alpha$ Col,
where MRI occurs in a shorter time. The reason for this is the fact that the wind speed for the $\alpha$ Col model is much faster, which causes greater shock heating of the protodisc. This is also the reason why there is a corresponding difference between the protodiscs of the B2 model and the model corresponding to $\alpha$ Ara in table \ref{table-protodisc-properties-bc-cohen}. Protodiscs of stellar models with the properties of $\alpha$ Ara and $\alpha$ Eri do not show any significant differences.

\begin{table}
\begin{minipage}{62mm}
\caption{Values of the equatorial density, height and time taken when a point $X$ becomes the outer edge in a protodisc of a uniformly rotating stellar model with the same parameters as $\alpha$ Ara with $\mathcal{S}_\star = 0.7$ and $B_{m, \circ}= 245$G.}
\label{table-protodisc-critical-ratios-alpha-ara-07-uniform}
 \begin{tabular}{rlrr}  \hline {$x$} & {$\rho_{X, edge}$}
 & {$h_{X, edge}/R$} & {$t_{X, edge}$} \\
  {} & {(gm $\textrm{cm}^{-3}$)}
 &  {} &  {(days)}\\
\hline
  $ 1.288$  & $ 1.10E-12$   & $0.156$   & $  29.836$\\
  $ 1.408$  & $ 2.03E-13$   & $0.164$   & $   8.691$\\
  $ 1.608$  & $ 4.21E-14$   & $0.185$   & $   3.502$\\
  $ 1.808$  & $ 1.34E-14$   & $0.208$   & $   1.978$\\
  $ 2.008$  & $ 5.24E-15$   & $0.230$   & $   1.278$\\
  $ 2.208$  & $ 2.32E-15$   & $0.251$   & $   0.889$\\
  $ 2.408$  & $ 1.13E-15$   & $0.270$   & $   0.648$\\
  $ 2.608$  & $ 5.87E-16$   & $0.288$   & $   0.489$\\
  $ 2.808$  & $ 3.23E-16$   & $0.302$   & $   0.378$\\
  $ 3.008$  & $ 1.86E-16$   & $0.314$   & $   0.297$\\
  $ 3.208$  & $ 1.11E-16$   & $0.322$   & $   0.237$\\
  $ 3.408$  & $ 6.90E-17$   & $0.325$   & $   0.190$\\
\hline
\end{tabular}

A necessary condition for the onset of MRI is not satisfied when the rotation is uniform.
\end{minipage}
\end{table}

\begin{table*}
\begin{minipage}{112mm}
\caption{Values of the equatorial density, height and time taken when a point $X$ becomes the outer edge or when MRI occurs at $X$ in a protodisc of a differentially rotating stellar model with the same parameters as $\alpha$ Ara with $\mathcal{S}_{\star, eq} =
0.7$, $\epsilon = 0.2$ and $B_{m, \circ}= 245$G.}
\label{table-protodisc-critical-ratios-alpha-ara-07-diff}
 \begin{tabular}{rllrrrr}  \hline {$x$} & {$\rho_{X, edge}$}  & {$\rho_{X, MRI}$} & {$h_{X, edge}/R$} & {$h_{X, MRI}/R$} & {$t_{X, edge}$} & {$t_{X, MRI}$} \\
  {} & {(gm $\textrm{cm}^{-3}$)} & {(gm $\textrm{cm}^{-3}$)} &  {} &  {} &  {(days)}  &
  {(days)} \\
\hline
  $ 1.330$  & .....  & $ 2.62E-13$  & .....  & $0.149$  & .....  & $   8.384$\\
  $ 1.410$  & .....  & $ 1.34E-13$  & .....  & $0.158$  & .....  & $   5.771$\\
  $ 1.610$  & .....  & $ 3.56E-14$  & .....  & $0.183$  & .....  & $   2.973$\\
  $ 1.810$  & .....  & $ 1.27E-14$  & .....  & $0.207$  & .....  & $   1.880$\\
  $ 2.010$  & .....  & $ 5.42E-15$  & .....  & $0.232$  & .....  & $   1.328$\\
  $ 2.210$  & .....  & $ 2.61E-15$  & .....  & $0.256$  & .....  & $   1.005$\\
  $ 2.410$  & .....  & $ 1.38E-15$  & .....  & $0.280$  & .....  & $   0.797$\\
  $ 2.450$  & .....  & $ 1.22E-15$  & .....  & $0.284$  & .....  & $   0.765$\\[2pt]
  $ 2.510$  & $ 1.01E-15$  & .....  & $0.290$  & .....  & $   0.704$  & .....\\
  $ 2.610$  & $ 7.35E-16$  & .....  & $0.300$  & .....  & $   0.617$  & .....\\
  $ 2.810$  & $ 4.08E-16$  & .....  & $0.318$  & .....  & $   0.481$  & .....\\
  $ 3.010$  & $ 2.37E-16$  & .....  & $0.333$  & .....  & $   0.383$  & .....\\
  $ 3.210$  & $ 1.43E-16$  & .....  & $0.346$  & .....  & $   0.309$  & .....\\
  $ 3.410$  & $ 8.92E-17$  & .....  & $0.355$  & .....  & $   0.252$  & .....\\
  $ 3.610$  & $ 5.72E-17$  & .....  & $0.361$  & .....  & $   0.207$  & .....\\
  $ 3.810$  & $ 3.77E-17$  & .....  & $0.362$  & .....  & $   0.171$  & .....\\
  $ 4.010$  & $ 2.54E-17$  & .....  & $0.357$  & .....  & $   0.141$  & .....\\
  $ 4.190$  & $ 1.81E-17$  & .....  & $0.347$  & .....  & $   0.118$  & .....\\
\hline
\end{tabular}
\end{minipage}
\end{table*}

\begin{table*}
\begin{minipage}{110mm}
\caption{Values of the equatorial density, height and time taken when a point $X$ becomes the outer edge or when MRI occurs at $X$ in a protodisc of a differentially rotating stellar model with the same parameters as $\alpha$ Ara with $\mathcal{S}_{\star, eq} =
0.5$, $\epsilon = 0.2$ and $B_{m, \circ}= 119$G.}
\label{table-protodisc-critical-ratios-alpha-ara-05-diff}
 \begin{tabular}{rllrrrr}  \hline {$x$} & {$\rho_{X, edge}$}  & {$\rho_{X, MRI}$} & {$h_{X, edge}/R$} & {$h_{X, MRI}/R$} & {$t_{X, edge}$} & {$t_{X, MRI}$} \\
  {} & {(gm $\textrm{cm}^{-3}$)} & {(gm $\textrm{cm}^{-3}$)} &  {} &  {} &  {(days)}  &
  {(days)} \\
\hline
  $ 1.70$  & .....  & $ 5.34E-15$  & .....  & $0.163$ & .....  & $   0.577$\\
  $ 2.00$  & .....  & $ 1.37E-15$  & .....  & $0.188$ & .....  & $   0.322$\\
  $ 2.40$  & .....  & $ 3.40E-16$  & .....  & $0.215$ & .....  & $   0.186$\\
  $ 2.80$  & .....  & $ 1.12E-16$  & .....  & $0.235$ & .....  & $   0.122$\\
  $ 3.20$  & .....  & $ 5.27E-17$  & .....  & $0.242$ & .....  & $   0.091$\\
  $ 3.60$  & .....  & $ 2.00E-17$  & .....  & $0.234$ & .....  & $   0.059$\\
  $ 4.00$  & .....  & $ 9.95E-18$  & .....  & $0.201$ & .....  & $   0.039$\\
  $ 4.20$  & .....  & $ 7.24E-18$  & .....  & $0.167$ & .....  & $   0.029$\\
  $ 4.24$  & .....  & $ 6.81E-18$  & .....  & $0.159$ & .....  & $   0.026$\\
  $ 4.26$  & .....  & $ 6.61E-18$  & .....  & $0.154$ & .....  & $   0.025$\\[2pt]
  $ 4.28$  & $ 6.39E-18$  & .....  & $0.148$  & .....  & $   0.024$  & .....\\
  $ 4.38$  & $ 5.35E-18$  & .....  & $0.099$  & .....  & $   0.015$  & .....\\
  $ 4.44$  & $ 4.83E-18$  & .....  & $0.043$  & .....  & $   0.006$  & .....\\
\hline
\end{tabular}
\end{minipage}
\end{table*}

Tables \ref{table-protodisc-critical-ratios-alpha-ara-07-uniform} shows critical values of properties of a protodisc when a point $X$ on the equatorial plane becomes the outer edge at $X$ for a uniformly rotating stellar model having the properties of $\alpha$ Ara with $\mathcal{S}_\star = 0.7$. Tables \ref{table-protodisc-critical-ratios-alpha-ara-07-diff} and \ref{table-protodisc-critical-ratios-alpha-ara-05-diff} show the corresponding values when $X$ becomes the outer edge or when MRI occurs at $X$ for differentially rotating models with $\mathcal{S}_{\star, eq} = 0.7$ and $\mathcal{S}_{\star, eq} = 0.5$, respectively, $\epsilon = 0.2$ and having the same basic
stellar properties.  Suppose that we consider the same basic model with different rotation rates and different magnetic field strengths. Then, when the rotation rate is fixed and the magnetic field strength is increased, the value of $x_2$ increases slightly. The results for stellar models with the properties of $\alpha$ Ara are shown in table \ref{table-protodisc-properties-alpha-ara-uni-diff-rot-mag} for uniform rotation and for differential rotation with $\epsilon = 0.2$. Also, when the stellar rotation rate is increased while the magnetic field strength is fixed, the value of $x_2$ decreases. We note that the values of $x_2$ for models with differential rotation are somewhat larger than those for the corresponding models with uniform rotation. Our results show that when stars
have solar-type differential rotation, there are broad ranges of stellar magnetic field strengths and rotation rates such that $\mathcal{S}_{\star, eq} \leq 0.9$ for which MRI sets in over a significant region from $X_{inner}$ to $X_2$  before the protodiscs shrink radially to that region. However, when a star rotates with near critical rotation speed such that $\mathcal{S}_\star > 0.9$, we find that the time for the onset of MRI is longer than the time for the protodisc region to shrink to a thin ring, unless some other MHD instability occurs.

\begin{table*}
\begin{minipage}{142mm}
\caption{Variation of properties of protodiscs with changes in
rotation speed and magnetic field strength for a stellar model
with the parameters of $\alpha$ Ara, dipole $b=3$ and
$\beta_{vel}=1$ for uniform and differential rotation.}
\label{table-protodisc-properties-alpha-ara-uni-diff-rot-mag}
\begin{tabular}{lcrrrrrrrrrrrc} \hline \multicolumn{3}{c}{} &\multicolumn{4}{c}{Uniform
Rotation with $\mathcal{S}_\star = \mathcal{S}_{\star, eq}$} & {}
&\multicolumn{5}{c}{Differential Rotation with $\epsilon = 0.2$}
\\ \cline{4-7} \cline{9-13}
\\
 {$\mathcal{S}_{\star, eq}$} & {$B_{m, \circ}$}
& {$x_{lim}$} & {$B_{\phi, \circ}$} & {$x_{inner}$} & {$x_1$} &
{$t_{shrink}$} & {} & {$B_{\phi, \circ}$} & {$x_{inner}$} &
{$x_1$} & {$x_2$} & {$t_2$}
\\
 {} & {(G)} & {(G)} & {} & {}
& {} & {(days)}  & {} & {(G)} & {} & {} & {} & {(days)} \\ \hline
 $0.40$  & $ 245$  & $ 9.63$  & $  -7.46$  & $ 1.84$  & $ 6.05$  & $6.19$ & & $  -6.13$  & $ 1.97$  & $ 6.62$  & $ 5.79$  & $   0.089$\\
 $0.50$  & $ 245$  & $ 9.63$  & $  -9.33$  & $ 1.59$  & $ 4.92$  & $ 10.61$ & & $  -7.66$  & $ 1.68$  & $ 5.81$  & $ 4.30$  & $   0.202$\\
$0.60$  & $ 245$  & $ 9.63$  & $ -11.20$  & $ 1.41$  & $ 4.16$  & $ 17.89$  & & $  -9.19$  & $ 1.47$  & $ 4.88$  & $ 3.27$  & $   0.378$\\
$0.70$  & $ 245$  & $ 9.63$  & $ -13.06$  & $ 1.27$  & $ 3.60$  & $ 29.84$ & & $ -10.72$  & $ 1.31$  & $ 4.20$  & $ 2.45$  & $   0.765$ \\
$0.80$  & $ 245$  & $ 9.63$  & $ -14.93$  & $ 1.16$  & $ 3.17$  & $ 47.54$ & & $ -12.25$  & $ 1.19$  & $ 3.68$  & $ 1.55$  & $   3.690$ \\
$0.90$  & $ 245$  & $ 9.63$  & $ -16.80$  & $ 1.07$  & $ 2.82$  & $ 65.38$ & & $ -13.79$  & $ 1.08$  & $ 3.27$  & $ 1.08$  & $ 70.030$ \\[{4pt}]
$0.75$  & $  57$  & $ 3.97$  & $  -6.43$  & $ 1.21$  & $ 2.40$  & $ 2.51$ & & $  -5.47$  & $ 1.24$  & $ 2.73$  & $ 1.84$  & $   0.114$ \\
$0.75$  & $ 119$  & $ 6.15$  & $  -9.34$  & $ 1.21$  & $ 2.86$  & $ 9.74$ & & $  -7.77$  & $ 1.24$  & $ 3.35$  & $ 1.94$  & $   0.387$ \\
$0.75$  & $ 245$  & $ 9.63$  & $ -14.00$  & $ 1.21$  & $ 3.38$  & $ 38.05$ & & $ -11.49$  & $ 1.24$  & $ 3.93$  & $ 2.04$  & $   1.287$ \\
$0.75$  & $ 499$  & $15.15$  & $ -21.43$  & $ 1.21$  & $ 3.98$  & $ 149.80$ & & $ -17.43$  & $ 1.24$  & $ 4.59$  & $ 2.12$  & $   4.453$ \\

\hline
\end{tabular}
\end{minipage}
\end{table*}

\section{Evolution of Disc Material after Onset of MRI}\label{sec-evln-protodisc-after-mri}
A detailed study of the evolution of disc material after the onset of MRI is beyond the goals of the present paper. However, we consider several factors that will play a role in the evolution from the protodisc phase to a quasi-steady disc. Because the electrical conductivity is extremely high in a protodisc region, the magnetic diffusivity is negligible and the Ohmic diffusion time is extremely long \citep{mah03}. Thus, before the onset of MRI, the magnetic field within a protodisc region remains frozen in the material as it rotates. At the outer edge, the sum of the centrifugal force and thermal pressure exceeds the sum of gravity and the magnetic force, so that disc material will flow outwards and draw out magnetic field lines with it. We focus on stellar models where magnetic fields are sufficiently strong to channel wind flow into disc regions. In the case of B type stars, field strengths of order 1G to 10G would be appropriate. We wish to understand how the onset of MRI in protodiscs of such stars helps in the formation of quasi-steady discs that survive over a period of several decades. Turbulence arising from MRI may persist for a long time and will cause changes in the magnetic field and the magnetic diffusivity of the disc material will be affected. The occurrence of MHD dynamo action is possible. A study of the evolution of the magnetic field will be complicated. In a quasi-steady state we expect that the structure and strength of the field will allow a slow outflow of the disc material.

After the onset of MRI, viscosity becomes important and the Shakura-Sunyaev viscosity coefficient is given by $\alpha \approx B^2/(4 \pi \rho \, c_s^2)$ subject
to the condition that $\alpha \leq 1$. The azimuthal viscous force is negative and will tend to decelerate the disc rotation speed. Also, in general, ${\partial B_\phi}/{\partial z}$ will no longer be zero and the azimuthal magnetic force $\mathcal{F}_\phi$ may play a role in the rotation of the disc material. The importance of viscosity has been invoked in the modelling of decretion discs \citep[e.g.,][]{lee91, oka01} and accretion discs \citep[e.g.,][]{pri81, bal03}. In decretion discs, material is fed into the disc region from the stellar equator and drifts outwards due to the effect of viscosity. In accretion discs, material in Keplerian discs fall towards the central star due to viscous loss of angular momentum. These studies consider thin discs and set up equations for physical quantities averaged over the heights of  discs. We have not used such an approach in our models of protodiscs. However, because the heights of protodiscs are relatively small at the time that MRI occurs, thin disc models may be a convenient alternative in a study of the evolution of disc material after the onset of MRI. \citet{haw95, haw96} have used a shearing box model for thin accretion discs and found that there will be an outward flux of angular momentum. We are interested in scenarios in which the angular velocity distribution of the disc material readjusts so that a quasi-steady disc may be formed, possibly with inflow and slow outflow. A model conforming to such a description has been developed by \citet{bro08} where wind material is fed into a disc region by a magnetic field and flows slowly outwards with quasi-Keplerian rotation speed. Another possible quasi-steady model may be a magnetic rotator decretion disc model in which a protodisc that suffers MRI will provide material with super-Keplerian rotation speed to a decretion disc region. Such a model will be more realistic than the decretion disc model studied by \citet{lee91} and \citet{oka01} because the nonmagnetic models of Lee et al and Okazaki require the extremely special condition that Keplerian rotation must occur at the stellar equator.

In the case of a quasi-steady disc, continuity of mass and magnetic flux require that $\partial(\varpi \rho u_\varpi)/\partial \varpi = 0$, $\textrm{div}\bvect{B} = 0$ and an appropriate MHD induction equation be satisfied. The radial component of the momentum equation for axially symmetric quasi-steady discs with $u_z = 0$ can be written as
\begin{equation}\label{eqn-qsdisk-radial}
u_\varpi \frac{\partial u_\varpi}{\partial \varpi} = \frac{u_\phi^2}{\varpi} - \frac{GM}{\varpi^2} - \frac{1}{\rho} \frac{\partial p}{\partial\varpi} + \frac{\mathcal{F}_\varpi}{\rho} \, ,
\end{equation}
where
\begin{equation}\label{eqn-magnetic-force-radial-comp}
\mathcal{F}_\varpi = \frac{1}{4 \pi} \left[B_z \left(\frac{\partial B_\varpi}{\partial z} - \frac{\partial B_z}{\partial \varpi}\right) - \frac{B_\phi}{\varpi}\frac{\partial}{\partial \varpi}\left(\varpi B_\phi\right)\right]
\end{equation}
is the radial component of the magnetic force. The thermal pressure term in equation (\ref{eqn-qsdisk-radial}) may be negligible in a quasi-steady state where the disc material has cooled down and a radiative force may be included when appropriate. We are interested in quasi-steady models in which the radial speed $u_\varpi$ is small and the rotation speed $u_\phi$ is approximately equal to the magnetically modified Keplerian speed $\hat{u}_\phi = [GM(1-\zeta)/\varpi]^{1/2}$ defined in \citet{mah03}, where $\zeta = \varpi^2 \mathcal{F}_\varpi/(GM\rho)$ is the ratio of the radial component of the magnetic force to gravity and is small. The azimuthal component of momentum satisfies
\begin{equation}\label{eqn-qsdisk-azimuthal}
\frac{u_\varpi}{\varpi} \frac{\partial}{\partial \varpi}\left(\varpi u_\phi\right) = \frac{\mathcal{F}_\phi}{\rho} + \frac{1}{4 \pi \rho \varpi^2} \frac{\partial}{\partial\varpi}\left(\varpi^2 B^2\right) \, .
\end{equation}
The last term in this equation represents the $\alpha$-viscosity force and will be negative. When $u_\varpi$ is positive, the excess angular momentum brought in by the wind material is removed by the outward drift of the disc material. We expect that $u_\varpi$ will be small as shown by \citet{bro08}. This is because the continuity of mass in a quasi-steady state requires that $\rho_W v_{n,W} = \rho_D u_{n,D}$ across the shock boundary. If the disc region is isothermal in a quasi-steady state, we have $\rho_D \approx \rho_W  v_{n,W}^2/c_s^2$ \citep{cas02} so that we obtain $u_\varpi \approx v_{n,W} (c_s^2/v_{n,W}^2)$. Because $c_s \ll v_{n,W}$ for the types of stars that we consider, we find that $u_\varpi \ll v_{n,W}$. The $z$-component of the momentum equation remains the same as (\ref{eqn-disc-proto-mom-z-0}). In a detailed study we will have to include appropriate jump conditions at the shock boundary of the disc.

At the inner end of a protodisc, if the azimuthal viscous force is sufficiently strong to decelerate the rotation speed, disc material will tend to flow back towards the star as in the case of an accretion disc. However, this can compress the field lines of a dipole type magnetic field causing an increase in the magnetic pressure that will oppose the backward flow. Also, continuing inflow of wind material will supply angular momentum and it may be possible for a quasi-steady state to exist for some length of time despite any backward flow that may occur at the inner boundary of the disc region. Obviously, a detailed study will be required to determine the nature of the dynamics in this region. Although there may be other types of MHD instability that could affect the angular velocity distribution in a protodisc, we focus on MRI because we are able to determine the conditions that must be satisfied by the central star for its occurrence in a protodisc. An important aspect of our study of protodiscs of differentially rotating stars is that scenarios may exist in which MRI can intervene to redistribute angular momentum of the disc region and, possibly, permit the formation of quasi-steady disc regions.

\section{Discussion}\label{sec-discussion}
In this paper, we set up equations and jump conditions to study the formation, structure and evolution of protodiscs around rotating magnetic stars with winds. We obtain solutions for a variety of stellar models with different rotation rates and magnetic field strengths. When the magnetic field of a rotating hot star is strong enough to channel the flow of wind material towards the equatorial plane and field lines from the stellar surface cross the equatorial plane and return to the star, a protodisc with super-Keplerian rotation and no radial motion will be formed in an equatorial region. Initially, the sum of the magnetic force and gravity in this region is sufficient to prevent any radial outflow that may be caused by the centrifugal force and thermal pressure. As wind material continues to flow into the protodisc, its equatorial density and height increase with time. The resulting increase in centrifugal force relative to the magnetic force causes the material at the outer edge of the protodisc to move outwards so that the radial extent of the protodisc region decreases. Shock heating causes protodisc temperatures rise to values of order $1 \, \textrm{to} \, 10 \times 10^6$K. Because the radiative cooling time is longer than the heating up time, protodiscs retain their high temperatures. There is no meridional flow of material within a protodisc and the continuing increase in density causes the thermal pressure in the protodisc to increase in relation to the magnetic pressure. When the rotation rate of the central star satisfies the condition $\mathcal{S}_{\star, eq} \leq 0.9$ and the angular velocity of the protodisc region decreases radially outwards, MRI will occur before the radial extent of the protodisc can decrease significantly. In the case of most stars with near critical rotation rates such that $\mathcal{S}_\star > 0.9$, the time taken for the onset of MRI is longer than the time by the protodisc to shrink to a thin ring. The evolution of protodiscs is strongly influenced by the inflow of wind material, centrifugal outflow and the occurrence of MRI. The protodisc phase of a disc region terminates either when it shrinks completely or when some instability such as MRI sets in. We derive conditions for centrifugal outflow at the outer edge of a protodisc and for the onset of MRI in a protodisc. We perform numerical computations to study the structure and properties of protodiscs of several different stellar models and to determine how the protodisc phase terminates. Our results show that there are situations involving different field strength domains that lead to the formation of protodiscs that suffer MRI soon after their formation or shrink to a thin ring when no instability intervenes.

Equation (\ref{eqn-proto-relation-Omega-omega}) gives the angular velocity of a protodisc in terms of that of the central star. When the central star has uniform rotation, the protodisc has approximately uniform rotation and if no instability intervenes, its radial extent continues to decrease. A protodisc region may take a long time to shrink completely and during most of this time it would appear like a ring. However, it is not clear whether steady uniform rotation is possible in the surface regions of rapidly rotating early type stars because large scale meridional circulations must occur in the radiative envelopes of such stars. Meridional circulation will transport angular momentum and disrupt the angular velocity distribution to cause differential rotation. Studies by \citet{mah68} showed that steady rotation may be possible in the interior of the radiative envelope with angular velocity constant along field lines and having about
$3$ percent variation from uniform rotation. However, because of the sensitivity of meridional circulation near the stellar surface to perturbations in angular velocity, steady uniform rotation may not possible in the surface region. Other efforts to develop
models of steady uniform or differential rotation in early type stars with magnetic fields have been discussed by \cite{str70} and \citet{tas00}. We are not aware of any conclusive theoretical or observational evidence that early type stars with fast rotation and magnetic fields will have steady unform rotation along the surface.

When the stellar surface has differential rotation with angular velocity increasing towards the equator, equation (\ref{eqn-proto-relation-Omega-omega}) shows the angular velocity
of the protodisc decreases radially outwards. In this case, MRI occurs in the protodisc within a few hours or days. We know that such rotation occurs in the Sun, where the differential rotation is maintained by meridional circulation and convection \citep[see
e.g.,][]{tas00}. Recently, \citet{mae08} have suggested that convection zones will be present near the surface of rotating massive stars. This situation is the same as in the surface regions of the Sun where convective motions and meridional circulation are present. Thus, we suggest that it is reasonable to expect that there are massive stars with solar-type differential rotation.

Our study of protodiscs was motivated in large part by the claim that discs will not form around magnetic rotator stars because of centrifugal breakout found by \citet{udd06} in numerical simulations of uniformly rotating stars. \citet{mah03} had previously pointed out that centrifugal outflow of protodisc material due to increased density could diffuse into Keplerian, especially when viscosity was present. In studying the evolutionary properties of protodiscs, we have focussed on centrifugal outflow and the onset of MRI. It is possible that there are other MHD instabilities that may also play a role in the evolution of protodiscs to quasi-steady discs. We have been able to show that under certain highly plausible conditions at the stellar surface, MRI can occur in protodiscs before they shrink significantly through centrifugal outflow. Protodiscs in which MRI occurs in an inner part before the outer edge shrinks to that part are of interest because viscosity arising from the onset of MRI may be able to redistribute the angular velocity of the disc so that material may diffuse into a quasi-steady disc region with slow outflow. Although, constructing detailed models of quasi-steady discs is beyond the scope our present work, we simply point out that there are such models that are plausible. In the first model, the disc magnetic field remains connected to the stellar field and a quasi-steady disc forms with inflow from the wind zone and slow radially outward flow with field lines stretching out with the flow \citep[e.g.,][]{bro08}. Another possible quasi-steady model is one in which a protodisc evolves after the onset of MRI into a magnetic rotator decretion disc \citep[analogous to the model of][]{lee91} with inflow of super-Keplerian wind material. Therefore, the centrifugal breakouts found in the numerical simulations of \citet{udd06} in uniformly rotating models does not mean that quasi-steady discs cannot form around differentially rotating magnetic rotator stars.

We have not addressed the question of how protodiscs may be detected observationally. Because the duration of the protodisc phase in most cases is rather short, being of the order of a few hours to many days, special observational efforts will be required during disc formation phases to infer their presence. In some special cases, protodiscs can survive for a few years in the form of rings.

\section*{Acknowledgments}
We gratefully acknowledge the following grants:
(MM) UWMC Summer Research Grant; (JPC) The NSF Center for Magnetic
Self Organization in Laboratory and Astrophysical Plasmas Grant. We are grateful to John Brown for valuable discussions and comments on the paper. We thank an anonymous referee for useful comments on the paper.

{}

\label{lastpage}

\end{document}